\documentclass[prd,twocolumn,showpacs,preprintnumbers,floats,floatfix,nofootinbib]{revtex4}

\usepackage{graphicx}
\usepackage{dcolumn}
\usepackage{amsfonts}
\usepackage{amssymb}

\begin{document}

\title{Gravitational Wave Recoil and Kick Processes in the Merger of Two Colliding Black Holes:
The Non Head-on Case}

\author{R. F. Aranha$^{1,2}$, I. Dami\~ao Soares$^{1}$ and E. V. Tonini$^{3}$}

\address{$^{1}$Centro Brasileiro de Pesquisas F\'isicas, Rio de Janeiro 22290-180, Brazil, email:ivano@cbpf.br\\
$^{2}$Center for Relativistic Astrophysics, Georgia Institute of Technology, Atlanta, GA 30332, USA, email:rafael.aranha@physics.gatech.edu\\
$^{3}$Instituto Federal do Esp\'irito Santo, Vit\'oria 29040-780, Brazil, email:tonini@cefetes.br.}

\date{\today}

\begin{abstract}
We examine numerically the process of gravitational wave recoil in the merger of two black holes
in non head-on collision, in the realm of Robinson-Trautman spacetimes. Characteristic initial
data for the system are constructed, and the presence of a common apparent horizon implies that the
evolution covers the post-merger phase up to the final configuration of the remnant black hole.
The net momentum flux carried out by gravitational waves and the associated impulses are evaluated.
Our analysis is based on the Bondi-Sachs conservation laws for the energy momentum of the system.
The net kick velocity $V_{k}$ imparted to the merged system by the total gravitational wave impulse
is also evaluated. Typically for a non head-on collision the net momentum flux carried out by gravitational waves
is nonzero for equal-mass colliding black holes. The distribution of $V_{k}$ as a function of the symmetric mass ratio $\eta$
is well fitted by a modified Fitchett $\eta$-scaling law, the additional parameter modifying the law being a measure
of the nonzero gravitational wave momentum flux for equal-mass initial black holes.
For an initial infalling velocity $v/c \simeq 0.462$ of the
colliding black holes, and incidence angle of collision $\rho_0=21^{o}$, we obtain a maximum
$V_{k} \sim 121~{\rm km/s}$ located at $\eta \simeq 0.226$. For initial equal-mass
black holes ($\eta=0.25$) we obtain $V_{k} \sim 107~{\rm km/s}$.
Based on the integrated Bondi-Sachs momentum conservation law we discuss a possible definition of the
center-of-mass velocity of the binary merged system and show that -- in an appropriate inertial frame --
it approaches asymptotically the net kick velocity, which is the velocity of the remnant black hole
in this inertial frame. For larger values of $v/c$ we obtain substantially larger values of the net kick velocity,
e.g., for $v/c \simeq 0.604$ a maximum $V_{k} \sim 610~{\rm km/s}$ is obtained.

\end{abstract}


\maketitle

\section{Introduction}

The collision and merger of two black holes is presently considered to be an important astrophysical
configuration where processes of generation and emission of gravitational waves take place (cf.\cite{pretorius}
and references therein). The radiative transfer involved in these processes, evaluated in the full nonlinear
regime of General Relativity, shows that gravitational waves extract mass, momentum and angular momentum of
the source, and may turn out to be fundamental for the astrophysics of the collapse of stars and the formation and population of
black holes in galaxies\cite{baker,komossa,jonker}.
\par In this vein we have recently examined numerically the gravitational wave production and related
radiative processes, as mass-energy and momentum extraction by gravitational waves, in a general collision
of two Schwarzschild black holes\cite{aranha11}, described in the realm of
non-axisymmetric Robinson-Trautman (RT) spacetimes\cite{rt}. Characteristic initial data
that represent two instantaneously boosted black holes in non head-on collisions were constructed for
the RT dynamics. The initial data already present a common apparent horizon so that the dynamics
covers the post-merger phase of the system. The present paper focuses on the processes of momentum extraction
by contemplating the time behavior of the momentum fluxes of the gravitational wave emitted and the
associated total impulses imparted to the binary merged system. Our treatment is based on the
Bondi-Sachs energy momentum conservation laws\cite{ bondi,sachs,sachs1}. The use of the Bondi-Sachs formulation
gives a clear and simple picture of the kick processes generated in the recoil of the
the binary merged system which are related to the maximum impulse due to the
gravitational wave flux in each regime characterized by the initial data.
\par The kick processes and associated recoils in the merger of two black holes have been investigated within
several approaches, all of them connected to binary black hole inspirals. Post-Newtonian approximations
(cf. \cite{blanchet1} and references therein) estimated the
kick velocity accumulated during the adiabatical inspiral of the system up to its innermost stable circular
orbit (ISCO) plus the kick velocity accumulated during the plunge phase (from the ISCO up to the horizon);
the plunge contribution dominates the recoil but the evaluation of the dynamics in the latter stage is the most uncertain.
A recent treatment using post-Newtonian calculations and incorporating ``close-limit approximation'' calculations is given in \cite{blanchet}.
The first fully numerical verification of the recoil in nonspinning black hole binaries
was reported by Baker et al.\cite{baker} for a mass ratio $\sim 0.667$ of the two black holes.
Later Gonz\'alez et al.\cite{gonzalez} undertook a more complete full numerical relativity examination
of kicks in the merger of nonspinning binaries by contemplating a larger parameter domain.
Recently Rezzola et al.\cite{rezzolla} obtained an important injective relation between
the kick velocities and the effective curvature parameter of the global apparent horizon,
in head-on collisions of two black holes in RT dynamics using
initial data derived in Refs. \cite{aranha1,aranhaT}.
Some features of a non head-on collision are distinct from those of
a head-on collision\cite{aranha1,aranha} or from the merger of inspiral nonspinning binaries,
mainly for large values of the initial mass ratio parameter and of the incidence angle of collision
of the two initial colliding black holes. One of them is the nonzero net momentum flux carried out by
gravitational waves in the case of equal-mass initial colliding black holes.
Our results will be compared with previous estimates of black hole kicks from full numerical relativity simulations
of binary black hole inspirals (cf. \cite{gonzalez} and references therein).
\par The presence of kick processes and the magnitude of the associated recoil velocities can have important consequences
for astrophysical scenarios, as the evolution and the population of massive black holes in galaxies or in the
intergalactic medium\cite{merritt,favata}. Observational evidence of black hole recoils have been
reported in \cite{komossa,jonker} and references therein.
\par RT spacetimes\cite{rt,kramer1} are asymptotically flat solutions of Einstein's vacuum equations
that describe the exterior gravitational field of a bounded system
radiating gravitational waves. In a suitable coordinate system the metric can be expressed as
{\small
\begin{eqnarray}
\label{eq1}
\nonumber
ds^2&=&\Big({\lambda(u,\theta,\phi)- \frac{2 m_{0}}{r}+ 2 r \frac{{K}_{u}}{K}}\Big) d u^2 +2du dr\\
&-&r^{2}K^{2}(u,\theta,\phi) (d \theta^{2}+\sin^{2}\theta d \varphi^{2}).
\end{eqnarray}}
where
{\small
\begin{eqnarray}
\lambda(u,\theta,\phi)=\frac{1}{K^2}-\frac{(K_{\theta}~ \sin \theta/K)_{\theta}}{K^2 \sin \theta}
+\frac{1}{\sin^2 \theta}\Big(\frac{K_{\phi}^2}{K^4}-\frac{K_{\phi \phi}}{K^3} \Big).
\label{eq2}
\end{eqnarray}}
The Einstein vacuum equations for (\ref{eq1}) result in
{\small
\begin{eqnarray}
\label{eq3}
-6 m_{0}\frac{{K}_u}{K}+\frac{1}{2 K^2}\Big(\frac{(\lambda_{\theta} \sin
\theta)_{\theta}}{ \sin \theta}+\frac{\lambda_{\phi \phi}}{ \sin^2 \theta}\Big)=0.
\end{eqnarray}}
\noindent In the above, subscripts $u$, $\theta$ and $\phi$ denote derivatives with respect to
$u$, $\theta$ and $\phi$, respectively. $m_0$ is the only dimensional parameter of the geometry, which
fixes the mass and length scales of the spacetime. Eq. (\ref{eq3}), denoted RT equation, governs the dynamics
of the system and allows to propagate the initial data $K(u_0,\theta,\phi)$, given in the characteristic surface $u=u_0$,
for times $u > u_0$. For sufficiently regular initial data RT spacetimes exist globally for all positive $u$
and converge asymptotically to the Schwarzschild metric as $u \rightarrow \infty$\cite{chrusciel}.
The field equations present two stationary solutions which will play an important role in our discussions:
(i) the Schwarzschild solution corresponding to
$K=K_0= {\rm const}$, with $\lambda=1/K_{0}^2$ and mass $M_{Schw}=m_{0} K_{0}^{3}$; and (ii)

{\small
\begin{equation}
K(\theta,\phi)=\frac{K_0}{\cosh \gamma+ ({\bf{n}} \cdot \hat{\bf{x}})\sinh \gamma},
\label{eq4}
\end{equation}}

\noindent where $\hat{\bf{x}}=(\sin \theta \cos \phi,\sin \theta \sin \phi, \cos \theta)$ is the unit
vector along an arbitrary direction ${\bf{x}}$, and ${\bf{n}}=(n_1,n_2,n_3)$ is a constant unit vector
satisfying $n_{1}^{2}+n_{2}^{2}+n_{3}^{2}=1$. Also $K_0$ and $\gamma$ are constants. We note
that (\ref{eq4}) yields $\lambda=1/K_{0}^{2}$, resulting in its stationary character.
This solution can be interpreted\cite{bondi} as a boosted black hole along the axis determined by the unit vector
${\bf{n}}$ with boost parameter $\gamma$, or equivalently, with velocity parameter $v=\tanh \gamma$.
The $K(\theta,\phi)$ function (\ref{eq4}), which depends on three parameters, is a K-transformation of the generalized
Bondi-Metzner-Sachs (BMS) group\cite{bondi,sachs1} and represents a general Lorentz boost of the BMS.
\par The Bondi mass function of this solution is $m(\theta,\phi)=m_{0}K^3(\theta,\phi)$, and the total
mass-energy of this gravitational configuration is given by the Bondi mass
{\small
\begin{eqnarray}
\label{eq5}
\nonumber
M_B&=&(1/4\pi)\int^{2\pi}_{0} d \phi \int^{\pi}_{0}  m(\theta,\phi)\sin \theta ~d\theta\\
&=&m_{0} K_{0}^{3}\cosh \gamma = m_{0} K_{0}^{3}/ \sqrt{1-v^2}.
\end{eqnarray}}
The interpretation of (\ref{eq4}) as a boosted black hole is relative to
the asymptotic Lorentz frame which is the rest frame of the black hole when $\gamma=0$.
This asymptotic inertial frame will be the one we refer to in Section 2, in the definition of
the Bondi-Sachs energy-momentum, and in Section 5 when the center-of-mass motion of the merged
system is discussed.
\par In the paper we use units such that $G=c=1$; $c$ is however restored after the definition of the kick velocity.
Except where explicitly stated, all the numerical results were done for $\gamma=0.5$. In our computational work we used $m_0=10$
but the results are given in terms of $u/m_0$. We should note that we can always set $m_0=1$ in the RT equation (\ref{eq3})
by the transformation $u \rightarrow {\tilde u}=u/m_0$.

\section{The Bondi-Sachs Four Momentum for RT Spacetimes, the news and the initial data}
\par Since RT spacetimes describe the asymptotically flat exterior gravitational field of a bounded system
radiating gravitational waves and the initial data of its dynamics is prescribed on null characteristic surfaces,
they can be included in the the realm of the Bondi-Sachs 2+2 formulation of gravitational waves in
General Relativity\cite{bondi,sachs,sachs1}. Consequently suitable expressions
for physical quantities to be used in the description of gravitational
wave emission processes, as the Bondi-Sachs four momentum  and its conservation laws, must be
properly derived. From the supplementary vacuum Einstein equations $R_{UU}=0$, $R_{U \Theta}=0$, and $R_{U \Phi}=0$ in the
$2+2$ Bondi-Sachs formulation\cite{bondi,sachs} (where $(U,R,\Theta,\Phi)$ are Bondi-Sachs coordinates),
we obtain\cite{bondi,sachs}
{\small
\begin{eqnarray}
\label{eq6}
\nonumber
&&\frac{\partial m(u,\theta,\phi)}{\partial u}= - K \Big( {c_u^{(1)}}^2 + {c_u^{(2)}}^2 \Big)
+ \frac{1}{2}\frac{\partial}{\partial u}
\Big[3 c_{\theta}^{(1)} \cot \theta \\&+& 4 c_{\phi}^{(2)} \frac{\cos \theta}{\sin^2 \theta} -2 c^{(1)} + c_{\theta \theta}^{(1)}+ \frac{2}{\sin \theta}
 c_{\theta \phi}^{(2)} - \frac{1}{\sin^2 \theta} c_{\phi \phi}^{(1)} \Big],
\end{eqnarray}}
where $m(u,\theta,\phi)$ is the Bondi mass function and $c_u^{(1)}(u,\theta,\phi)$ and $c_u^{(2)}(u,\theta,\phi)$ are the two
{\it news} functions, in RT coordinates. The extra factor $K$ in the first term of the second-hand-side
of Eq. (\ref{eq6}) comes from the transformation of the Bondi time coordinate $U$ to the RT coordinate $u$,~${\rm lim}_{r \rightarrow \infty}=\partial U/\partial u=1/K$~\cite{kramer}.
For the RT spacetimes (\ref{eq1}), the {\it news} are expressed as\cite{cornish,aranhaN}
{\small
\begin{eqnarray}
\label{eq7}
\nonumber
c_{u}^{(1)}(u,\theta,\phi)&=&\frac{1}{2} \Big( {\mathcal P}_{\theta \theta}-{\mathcal P}_{\theta} \cot \theta -\frac{{\mathcal P}_{\phi\phi}}{\sin^2 \theta}\Big),\\
c_{u}^{(2)}(u,\theta,\phi)&=&\frac{1}{\sin \theta} \Big({\mathcal P}_{\theta \phi}- {\mathcal P}_{\phi} \cot \theta\Big),
\end{eqnarray}}
where we have introduced the variable ${\mathcal P} \equiv 1/K$, for notation convenience.
We remark that $c^{(1)}_{u}=0=c^{(2)}_{u}$ for the boosted Schwarzschild solution (\ref{eq4}), as should be expected.
Our derivation\cite{aranhaN} of Eqs. (\ref{eq7}) extends the work of von der G\"onna and Kramer\cite{kramer} to the case of non-axisymmetric
RT spacetimes, and agrees with Eq. (45) of Cornish and Micklewright\cite{cornish} modulo the use of the
RT time derivative and of the function ${\mathcal P}$ adopted in their representation of RT metric.
\par The Bondi-Sachs four momentum is defined as\cite{sachs1}
{\small
\begin{eqnarray}
\label{eq9}
P^{\mu}(u)= \frac{1}{4 \pi} \int^{2 \pi}_{0} d \phi \int^{\pi}_{0} m(u,\theta,\phi)~l^{\mu} \sin \theta~ d \theta,
\end{eqnarray}}
\noindent where $l^{\mu}=(1, -\sin \theta \cos \phi,-\sin \theta \sin \phi, -\cos \theta)$, relative to an
asymptotic Lorentz frame\footnote{The four vector $l^{\mu}$ actually defines the generators ($l^{\mu} \partial/\partial U $) of the BMS translations in the
temporal and Cartesian $x,y,z$ directions of the asymptotic Lorentz frame\cite{sachs1}.}. From Eq. (\ref{eq6}) the Bondi-Sachs four momentum conservation law follows
{\small
\begin{eqnarray}
\label{eq12}
\frac{d P^{\mu}(u)}{d u}= P_{W}^{\mu}(u),
\end{eqnarray}}
where
{\small
\begin{eqnarray}
\label{eq11}
P_{W}^{\mu}(u)= -\frac{1}{4 \pi} \int^{2 \pi}_{0} d \phi \int^{\pi}_{0} K~l^{\mu} \Big( {c_u^{(1)}}^2 + {c_u^{(2)}}^2 \Big) \sin \theta~ d \theta~~
\end{eqnarray}}
is the net flux of energy-momentum carried out by the the gravitational waves emitted.
We note that the term between square brackets in the right-hand-side of (\ref{eq6}) vanishes in the integrations
due to the boundary conditions satisfied by the {\it news}, $c^{(1)}=c^{(2)}=0$ and
$c^{(1)}_{\theta}=c^{(2)}_{\theta}=0$ at $\theta=0, \pi$.
The mass-energy conservation law (Eq. (\ref{eq12}) for $\mu=0$) is the Bondi mass formula. Our main interest in the present
paper is the analysis of the linear momentum conservation law (Eq. (\ref{eq12}) for $\mu=x,y,z$).
As we have discussed in \cite{aranha11} a general collision of two black holes in RT dynamics
is planar, namely, the dynamics is restricted to the plane determined by the directions of the two
initial colliding black holes. Therefore, without  loss of generality, we fix this plane as the
$(x,z)$-plane so that the momentum conservation equations relevant to our discussion reduce to
{\small
\begin{eqnarray}
\label{eq13}
\frac{d{\bf P}(u)}{d u}={\bf P}_{W}(u),
\end{eqnarray}}
where ${\bf P}_W(u)=(P^{x}_{W}(u),0,P^{z}_{W}(u))$, with
{\small
\begin{eqnarray}
\label{eq15-i}
{P}_{W}^{x}(u)&=&\frac{1}{4 \pi}  \int^{2 \pi}_{0}  d \phi \int^{\pi}_{0} \sin^2 \theta \cos \phi~ K \Big( {c_u^{(1)}}^2 + {c_u^{(2)}}^2 \Big) d \theta,~~~~~~\\
\label{eq15-ii}
{P}_{W}^{z}(u)&=&\frac{1}{4 \pi}  \int^{2 \pi}_{0} d \phi \int^{\pi}_{0} \cos \theta \sin \theta~ K \Big( {c_u^{(1)}}^2 + {c_u^{(2)}}^2 \Big) d \theta.~~~~\\
\nonumber
\end{eqnarray}}
Obviously $P^{y}(u)$ is conserved, this fact being a consequence of $P_{W}^{y}(u)=0$ for all $u$.
\par The initial data to be used was derived in Ref. \cite{aranha11} and represents two instantaneously
colliding Schwarzchild black holes in the $(x,z)$ plane, at $u=u_0$,
{\small
\begin{eqnarray}
\nonumber
&&K(u_0,\theta,\phi)=\Big(\frac{1}{\sqrt{{\cosh \gamma+ \cos \theta \sinh \gamma}}}+\\
&&\frac{\alpha}{\sqrt{{\cosh \gamma - (\cos \rho_0 ~\cos \theta+ \sin \rho_0~ \sin\theta \cos \phi)\sinh \gamma}}}\Big)^2.~~~~~
\label{eq16}
\end{eqnarray}}
For $\alpha=0$, Eq. (\ref{eq16}) corresponds to a black hole boosted along the positive $z$-axis. The parameter $\rho_0$ of the
initial data, denoted the incidence angle of collision, defines the direction of the second initial black hole with respect to the $z$-axis;
$\alpha$ is the mass ratio of the two initial colliding black holes and $v=\tanh \gamma$ their initial infalling velocity.
This data already has a common apparent horizon so that the evolution covers the post merger regime up to the final configuration,
when the gravitational wave emission ceases.
Finally we note that, in the full Bondi-Sachs problem, further data -- the
{\it news} functions -- are needed to determine the evolution of the system. However for the RT dynamics the {\it news}
are already specified once the data (\ref{eq16}) is given, cf. Eqs. (\ref{eq7}).
\section{Numerical evolution}
The initial data (\ref{eq16}) is evolved numerically via the RT equation, which is integrated using a
Galerkin spectral method with a projection basis space in two variables (the spherical harmonics)\cite{fletcher}.
The numerical method is accurate and highly stable to long time runs so that we are able to reach the
final configuration of a boosted remnant black hole when the gravitational wave emission ceases\cite{aranha11}.
This long time computation is demanded in order to determine the behavior of basic quantities necessary for the
evaluation of the processes as, for instance, the total impulse imparted to the merged black hole
by the gravitational radiation. Therefore we can examine physical phenomena in the nonlinear regime
where full numerical relativity simulations might present some difficulties due, for instance, to the
limited computational domain. In this sense the results of the evolution of the above data may
be considered as complementary to full numerical relativity simulations on describing the late post-merger regime.
\par In our numerical experiments in the present paper we vary $\alpha$ in the interval $[0,1]$
with fixed $\rho_0=21^{o}$ and $\gamma=0.5$ (corresponding to an initial infalling velocity $v \simeq 0.462$).
For comparison purposes, results for other values of the initial parameter $\rho_0$ are also considered.
The truncation of the method adopted in our computation is $N_P=7$. Exhaustive numerical experiments show that
after a sufficiently long time $u \sim u_f$ all modal coefficients of the
Galerkin expansion become constant up to twelve significant digits, corresponding to a final time of computation $u_{f}$.
At $u_f$ the gravitational wave emission is considered to effectively cease.
From the final constant modal coefficients we reconstruct $K(u_f,\theta,\phi)$ that, in all cases, can be approximated as
{\small
\begin{eqnarray}
\label{eq17}
K(u_f,\theta,\phi) \simeq \frac{K_{f}}{\cosh \gamma_{f}+ (n_{1f} \sin \theta \cos \phi+n_{3f} \cos \theta) \sinh \gamma_{f}}.~~
\end{eqnarray}}
With the final parameters $(K_f,\gamma_f,n_{1f},n_{3f})$ obtained from the final modal coefficients, we have in all cases
that the rms error of Eq. (\ref{eq17}) is of the order of, or smaller than $10^{-12}$.
The final configuration corresponds then to a boosted Schwarzschild black hole
(cf. (\ref{eq4})) with a final velocity $ v_{f}={\rm tanh} \gamma_{f}$
along the direction determined by ${\bf{n}}_f=(n_{1f},0,n_{3f})$, and a final Bondi rest mass $m_0 K_{f}^{3}$.
In general $\gamma_{f}< \gamma$ and $K_f>K_0$. The angle $\rho_f=\arccos~(n_{3f})$ -- denoted the scattering
angle -- defines the direction of the remnant with respect to the $z$-axis. Within the numerical error of our
computation we have $(n_{1f})^2+(n_{3f})^2=1$, as expected.
\par The values of the parameters of the remnant black hole are one of the basic results
to be extracted from our numerical experiments that, together with the initial data,
allow us to evaluate quantities which are characteristic of the radiative transfer processes involved
in the gravitational wave emission. Also, by computing $K(u,\theta,\phi)$ for all $u>u_0$, we can obtain
the time behavior of important physical quantities, as for instance the total impulse imparted to the merged
binary system by the emission of gravitational waves. Our numerical results are displayed in Table 1 and constitute
the basis of the analysis of the gravitational wave recoil, in the next Section.
\section{Gravitational wave momentum fluxes and the kick processes in a non head-on collision}

We are now ready to examine the processes of momentum extraction and the associated impulses imparted
to the binary merged system by the emission of gravitational waves. Our starting point is the construction,
via the numerically integrated function $K(u,\theta,\phi)$, of the curves of the net momentum fluxes
carried out by gravitational waves.
\par In Fig. \ref{fig1} we display the curves of the net momentum fluxes $P_{W}^{x}(u)$ and $P_{W}^{z}(u)$
for mass ratios $\alpha=0.2,~0.5$, and fixed $\gamma=0.5$ and incidence angle $\rho_0=21^{o}$. In both cases the
curves present an initial regime of positive flux $P_{W}^{z}(u)$ up to $u=u_k$ when $P_{W}^{z}(u_k)=0$, with consequent increase of
the linear momentum of the merged system along this direction, $d P^{z}(u)/d u >0$.
The value of $u_k$ increases as $\alpha$ increases.
\par The net momentum flux $P_{W}^{x}(u) <0$ for all $u_0< u \leq u_f$, so that the linear momentum
of the system along the $x$ direction always decreases. We then see that the system undergoes a
dominant deceleration regime until it reaches the final configuration, the remnant boosted Schwarzschild black hole.
\par The initial regime of positive flux $P_{W}^{z}(u)>0$ (for $u_0 < u < u_k$) is not present in the case of larger initial data
parameters $\alpha$ and/or $\rho_0$. In the next Section we will discuss the role of this initial positive gravitational wave flux
in engendering, when present, an inspiral branch in the motion of the center-of-mass of the merged system.
\subsection{The integrated fluxes of gravitational waves and the total impulse imparted to the merged system}
\par Integrating in time the conservation Eq. (\ref{eq13}) we find that
{\small
\begin{eqnarray}
\label{eq18}
{\bf P}(u)- {\bf P}(u_0)={\bf I}_{W}(u),
\end{eqnarray}}
where ${\bf I}_{W}(u)=({I}_{W}^{x}(u),0,{I}_{W}^{z}(u))$
is the impulse imparted to the binary merged system due to the momentum carried out by the gravitational waves
emitted up to the time $u$. In the above
{\small
\begin{eqnarray}
\label{eq20-i}
\nonumber
{I}_{W}^{x}(u)&=&\frac{1}{4 \pi} \int^{u}_{u_0} d u^{\prime}  \int^{2 \pi}_{0} d \phi \times \\
&&\int^{\pi}_{0} \sin^2 \theta \cos \phi~ K \Big( {c_{u^{\prime}}^{(1)}}^2 + {c_{u^{\prime}}^{(2)}}^2 \Big) d \theta,
\end{eqnarray}}
and
{\small
\begin{eqnarray}
\label{eq20-ii}
\nonumber
{I}_{W}^{z}(u)&=&\frac{1}{4 \pi} \int^{u}_{u_0}d u^{\prime} \int^{2 \pi}_{0} d \phi \times \\
&& \int^{\pi}_{0} \cos \theta \sin \theta~ K \Big( {c_{u^{\prime}}^{(1)}}^2 + {c_{u^{\prime}}^{(2)}}^2 \Big) d \theta.
\end{eqnarray}}
The net total impulse imparted to the system has a dominant contribution from the deceleration regimes
(where $P_{W}^{z}(u)<0$ and $P_{W}^{x}(u)<0$) and will correspond to a net kick on the merged system.
This physically important behavior has already been reported in accurate numerical relativity simulations
of the merger of unequal-mass binary black hole inspirals\cite{blanchet,baker,gonzalez}.
\begin{figure}
\begin{center}
{\includegraphics*[height=7.0cm,width=9.1cm]{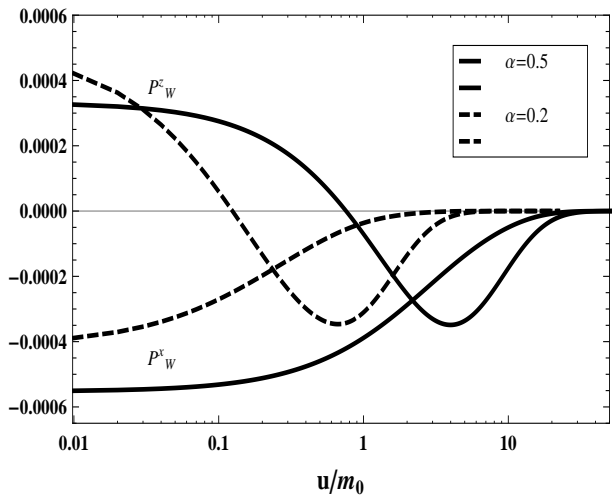}}
\caption{Linear-log plot of the net fluxes of momentum $P_{W}^{x}(u)$ and $P_{W}^{z}(u)$ for $\alpha=0.2,~0.5$,
and fixed $\gamma=0.5$ and incidence angle $\rho_0=21^{o}$. For both cases, an initial regime of positive
momentum flux along the $z$ axis is present, corresponding to an increase of the $z$-component of the
linear momentum, $d P^{z}(u)/du >0$, up to $u=u_k$ when $P_{W}^{z}(u_k)=0$. The dominant contribution to the net
impulse imparted to the system comes from the deceleration phases and will correspond to a net kick on the merged system.}
\label{fig1}
\end{center}
\end{figure}
\begin{figure}
\vspace{0.4cm}
\begin{center}
{\includegraphics*[height=6.5cm,width=8.1cm]{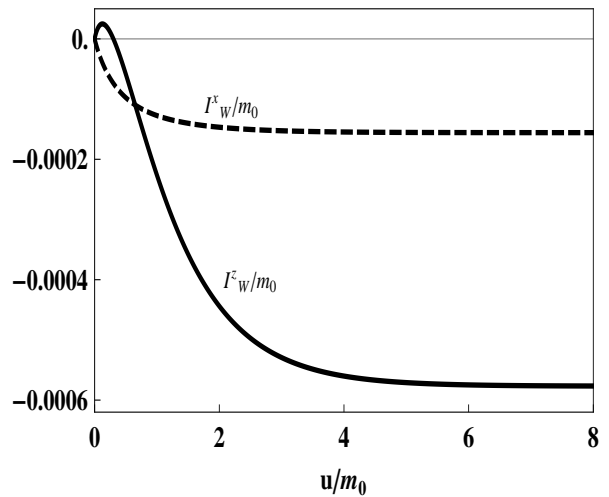}}
\caption{Plot of the impulses of the gravitational waves $I_{W}^{x}(u)$ (dashed curve) and $I_{W}^{z}(u)$
(continuous curve), for $\alpha=0.2$ and fixed initial incidence angle and infalling velocity, $\rho_0=21^{o}$
and $v \simeq 0.462$, respectively. The curve of $I_{W}^{z}(u)$ presents a local maximum at $u=u_k$, when the
deceleration regime along $z$ starts. At $u=u_m$, $I_{W}^{z}(u_m)=0$. For large $u \sim u_f$ both curves
tend to a constant negative value, corresponding to the final configuration when the gravitational wave emission ceases.}
\label{fig2}
\end{center}
\end{figure}
\begin{center}
\begin{table*}[t]
\caption{Summary of our numerical results corresponding to an infalling velocity
$v/c \simeq 0.4621$ ($\gamma=0.5$) and incidence angle of collision $\rho_0=21^{o}$.
The best fit of the points ($V_{k},\eta$) to the formula (\ref{eq24}) corresponds to
$A\simeq 3.63538199$, $B\simeq 2.46152380$ and $C \simeq 0.91379025$,
(cf. Fig. \ref{fitchett}).
}
\vspace{0.15cm}
\begin{tabular}{|c|c|c|c|c|c|c|c|c|c|c|c|}
\hline
$\alpha$ & $\eta$ & $K_f$ &$v_{in}/c$ &$v_f/c$&$\delta V/c$ & $\rho_f$ & $-I_{W}^{x}(u_f)/m_0$ & $-V_{k}^{x}$ & $- I_{W}^{z}(u_f)/m_0$ &$-V_{k}^{z}$& $V_{k}~({\rm km/s})$   \\\hline
$0.025$ & $0.0238$ &$1.045113$ &$0.446421$ &$0.446426$&1.5271 &$0.37^{o}$ & $1.2792 \times 10^{-6}$&0.3362 &$8.1563\times 10^{-6}$&$2.1435$ & $2.17$ \\ \hline
$0.050$ & $0.0453$ &$1.091454$ &$0.430848$ &$0.430867$&5.5605 &$0.77^{o}$ & $5.6819 \times 10^{-6}$&$1.3110$ &$3.3361\times 10^{-5}$ &$7.6973$& $7.81$ \\ \hline
$0.100 $ & $0.0826$ &$1.187836$&$0.400195$ &$0.400255$&17.9815&$1.65^{o}$ & $2.7635 \times 10^{-5}$&$4.9467$ &$1.3834 \times 10^{-4}$&$24.7624$&$25.25$ \\ \hline
$0.150$& $0.1134$ & $1.289162$ &$0.370369$&$0.370480$&34.0082&$2.64^{o}$ & $7.4349 \times 10^{-5} $&$10.4105$&$3.1932 \times 10^{-4}$&$44.7116$ & $45.91$ \\ \hline
$0.200$&$0.1388$& $1.395446$&$0.341541$ &$0.341701$&49.3652&$3.77^{o}$ &$1.5575 \times 10^{-4}$ & $17.1953$& $5.7681 \times 10^{-4}$& $63.6818$& $65.96$  \\ \hline
$0.250$ &$0.1600$& $1.506701$&$0.313841$ &$0.314042$&62.6056&$5.06^{o}$& $2.8314 \times 10^{-4}$& $24.8334$ & $9.0763\times 10^{-4}$& $79.6062$& $83.39$  \\ \hline
$0.300$& $0.1775$&$1.622933$&$0.287368$ &$0.287600$&72.7560&$6.53^{o}$& $4.6910 \times 10^{-4}$ & $32.9220$& $1.3051 \times 10^{-3}$ & $91.5906$& $97.33$ \\ \hline
$0.400$& $0.2040$& $1.870356$&$0.238376$&$0.238634$&82.7989&$10.17^{o}$& $1.0733 \times 10^{-3}$&$49.2111$ & $2.2569 \times 10^{-3}$&$103.4813$& $114.59$ \\ \hline
$0.500$& $0.2222$& $2.137755$&$0.194976$ &$0.195219$&80.6673&$15.06^{o}$& $2.0929 \times 10^{-3}$&$64.2673$&$3.3174 \times 10^{-3}$&$101.8707$ & $120.45$  \\ \hline
$0.525$& $0.2257$& $2.207669$&$0.185049$ &$0.185282$&79.1673&$16.53^{o}$& $2.4287 \times 10^{-3}$&$67.71537$&$3.5817 \times 10^{-3}$&$99.8653$ & $120.66$  \\ \hline
$0.600$& $0.2344$& $2.425151$&$0.157534$&$0.157731$&69.8608&$21.77^{o}$& $3.6700 \times 10^{-3}$&$77.1919$ & $4.3295 \times 10^{-3}$&$91.0621$& $119.38$  \\ \hline
$0.700$ &$0.2422$& $2.732560$&$0.126622$&$0.126756$&53.9660&$31.14^{o}$& $5.9634 \times 10^{-3}$&$87.6814$ &$5.0986 \times 10^{-3}$ & $74.9656$& $115.36$ \\ \hline
$0.800$& $0.2469$ &$3.059989$&$0.103262$&$0.103331$&35.8323&$44.09^{o}$ &$9.1477 \times 10^{-3}$&$95.7801$ &$5.3968 \times 10^{-3}$ &$56.5069$ & $111.21$ \\ \hline
$0.900$&$0.2493$ &$3.407441$&$0.088861$&$0.088880$&1.5271&$60.85^{o}$& $1.34131 \times 10^{-2}$&$101.7102$& $4.9651 \times 10^{-3}$& $37.6501$ & $108.45$ \\ \hline
$1.000$ &$0.2500$& $3.774920$&$0.084214$&$0.084214$&0.0016&$79.50^{o}$& $1.8965 \times 10^{-2}$&$105.7664$& $3.5148 \times 10^{-3}$& $19.6022$ & $107.57$ \\ \hline
\end{tabular}
\end{table*}
\end{center}
\par The behavior of the impulse is illustrated in Fig. \ref{fig2} where we plot $I_{W}^{x}(u)$ and $I_{W}^{z}(u)$ for $\alpha=0.2$,
$\rho_0=21^{o}$ and $\gamma=0.5$, corresponding to the time integration of the dashed curves in Fig. \ref{fig1}.
As expected the curve for $I_{W}^{z}(u)$ presents a local maximum at $u=u_k$, when the area measured below the curve of
$P_{W}^{z}(u)$ starts to give a negative contribution to the impulse, until $u=u_m$ when $I_{W}^{z}(u_m)=0$. For large
$u \sim u_f$ the curves tend to a constant negative value (a plateau), corresponding to the final configuration
when the gravitational wave emission ceases. It is worth mentioning that Fig. \ref{fig2} also illustrates
the dominance of a deceleration regime in the processes. The plateau is considered to be reached when
$|{\bf I}_{W}(u)-{\bf I}_{W}(u+h)| \lesssim 10^{-10}$,
where $h$ is the stepsize of the integration used for the evaluation of ${\bf I}_{W}(u)$.
At this stage the remnant black hole has a momentum ${\bf P}=(n_{1f},0,n_{3f})~P_f$, with
{\small
\begin{eqnarray}
\label{momentumF}
P_f=\sqrt{(P^{x}(u_f))^2+(P^{z}(u_f))^2}=m_0 K_{f}^{3} \sinh \gamma_f,
\end{eqnarray}}
whose distribution as function of $\alpha$, for several $\rho_0$, is given in \cite{aranha11}.
\par From Eq. (\ref{eq18}) we can derive that
{\small
\begin{eqnarray}
\label{eq21}
P^{x}(u_f)-P^{x}(u_0)&=&I^{x}_{W}(u_f),
\end{eqnarray}
\small
\begin{eqnarray}
\label{eq22}
P^{z}(u_f)- P^{z}(u_0) &=&I^{z}_{W}(u_f),
\end{eqnarray}}
where the right-hand sides of (\ref{eq21}) and (\ref{eq22}) are the nonzero components of the net total impulse
${\bf I}_W(u_f)$ generated by the gravitational waves emitted.

\par
\begin{figure}
\begin{center}
{\includegraphics*[height=7.0cm,width=9.1cm]{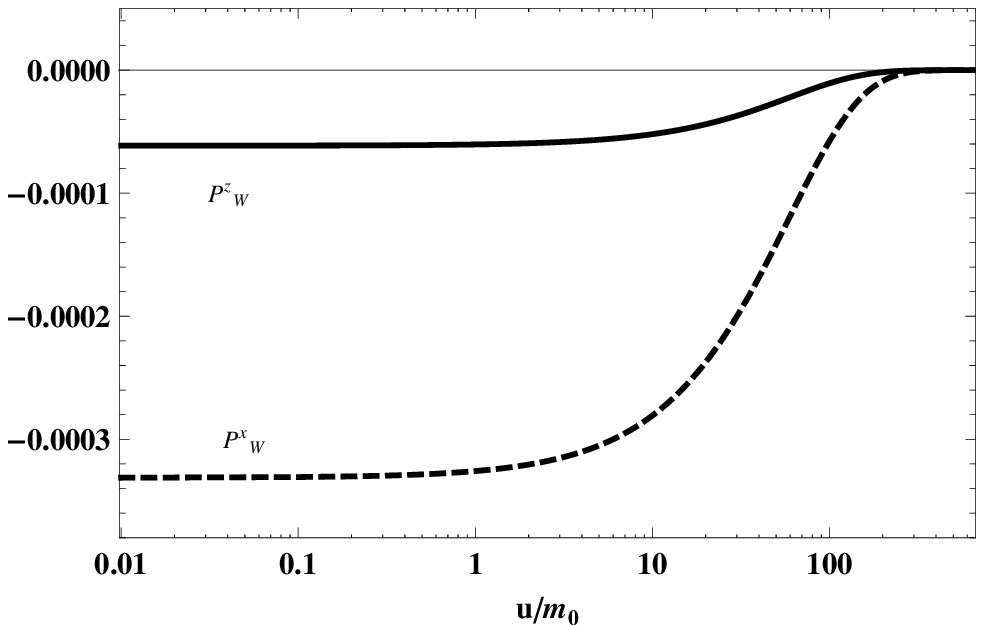}}
{\includegraphics*[height=7.0cm,width=9.1cm]{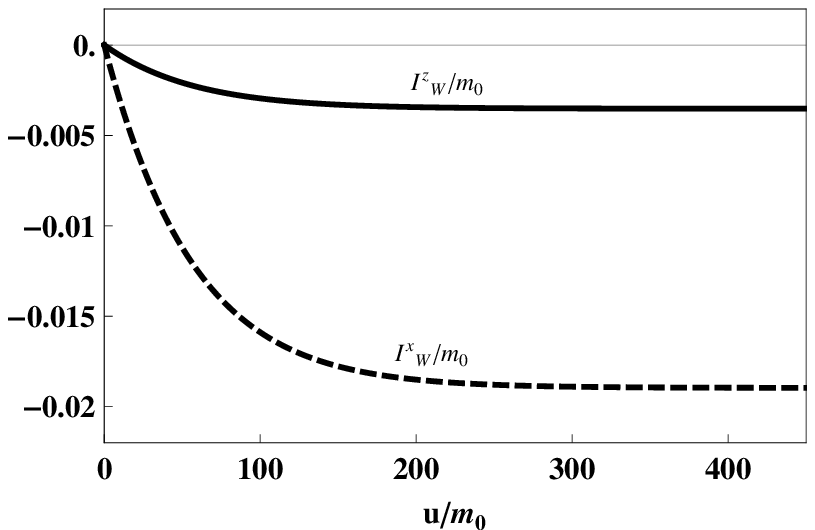}}
\caption{Linear-log plot of the net fluxes of momentum $P_{W}^{x}(u)$ and $P_{W}^{z}(u)$ (top)
and the associated total impulses $I_{W}^{x}(u)$ and $I_{W}^{z}(u)$ (bottom), for the case of equal-mass initial
colliding black holes ($\alpha=1$) and incidence angle $\rho_0=21^{o}$. The kick on the merged system for
this case is $V_{k} \sim 107~{\rm km/s}$.}
\label{fig3}
\end{center}
\end{figure}

 We remark that for larger values of $\alpha$, the final impulse $I_{W}^{x}(u_f)$ is larger than $I_{W}^{z}(u_f)$,
contrary to the cases $\alpha=0.2,~0.5$ shown in Fig. \ref{fig2}. This is shown in Figs. \ref{fig3} where we plot
the net momentum fluxes $P_{W}^{x}(u)$ and $P_{W}^{z}(u)$, and the associated impulses (integrated fluxes)
$I_{W}^{x}(u)$ and $I_{W}^{z}(u)$, for the case of equal-mass initial colliding black holes ($\alpha=1$)
and incidence angle $\rho_0=21^{o}$.
\par The purpose of Figs. \ref{fig3} is three-fold. First to show that, for high values of the
mass ratio, the initial phase of positive gravitational wave flux along the $z$-axis is absent, with an
overall deceleration of the binary merged system during the whole regime of gravitational wave emission.
Second, the total final impulse along the $z$ axis is about one order of magnitude smaller than the
total impulse along the $x$ axis; both of them are at least one order of magnitude larger that the corresponding
impulses for $\alpha=0.2$. Third, in a non-head-on collision of two equal-mass colliding
black holes, the net gravitational wave flux, and the associated impulses are nonzero, contrary to the cases of head-on collisions
or inspiral binaries of equal-mass black holes. The implications of the third point will be detailed in the next subsection.
\subsection{The kick velocity and the modified Fitchett distribution for non head-on collisions}

We are now led to define the net kick velocity ${\bf V}_{k}$ of the binary merged system as proportional to the
net momentum imparted to the system by the total impulse of gravitational waves up to $u=u_f$.
We remark that our definition of the net kick velocity is
based on the impulse function ${\bf I}_{W}(u)$ of the gravitational wave emitted
evaluated at $u=u_f$ (cf. Eqs. (\ref{eq21})-(\ref{eq22})). These definitions are in
accordance with \cite{gonzalez}. We obtain (restoring universal constants)
{\small
\begin{eqnarray}
\label{eq23-0}
{\bf V}_{k}&=&\frac{c}{m_0 K_{f}^{3}}~{\bf I}_{W}(u_f),
\end{eqnarray}}
with modulus
{\small
\begin{eqnarray}
\label{eq23}
\nonumber
V_{k}&=& \sqrt{(V_{k}^{x})^2+(V_{k}^{z})^2}\\
&=&\frac{c}{m_0 K_{f}^{3}}~\sqrt{I_{W}^{x}(u_f)^2+I_{W}^{z}(u_f)^2} ~,
\end{eqnarray}}
where $m_0 K_{f}^{3}$ is the rest mass of the remnant black hole.
We note that ${\bf I}_{W}(u_0)=0$. The velocity (\ref{eq23}) is directed along an axis
making the angle $\Theta_f=\arctan(I_{W}^{x}(u_f)/I_{W}^{z}(u_f))$ with the negative $z$-axis.
The interpretation of this angle, as defining the asymptotic direction of the
center-of mass motion in a particular inertial frame, will be given in Section V.
From our numerical results we evaluate $V_{k}$ for several values of $\alpha$, and for
fixed $\gamma=0.5$ and incidence angle $\rho_0=21^{o}$. The results are summarized in Table 1
where, in accordance with the literature, we use the symmetric mass parameter $\eta=\alpha/(1+\alpha)^2$.
\par
\begin{figure}
\begin{center}
{\includegraphics*[height=6.2cm,width=8.2cm]{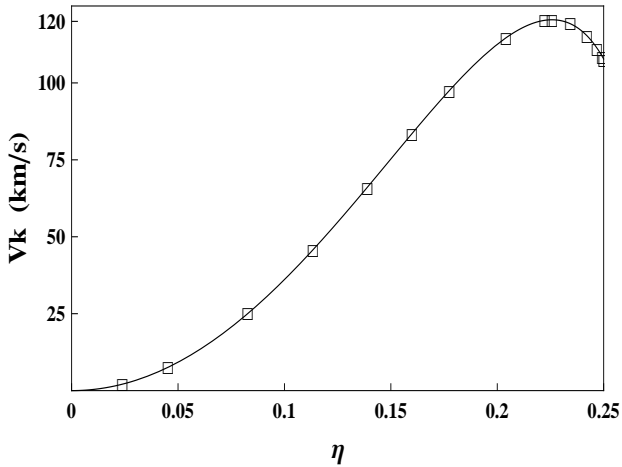}}
\caption{Plot of the points ($V_{k},\eta$), where $V_{k}$ is the net kick velocity
due to the impulse imparted to the merged system by the gravitational waves, for $\rho_0=21^{o}$.
The continuous curve is the least-square-fit of the points to the analytical formula (\ref{eq24}),
with best fit parameters $A\simeq 3.635382$, $B\simeq 2.461524$ and $C \simeq 0.913790$ (cf. Table 1).
The maximum of the curve corresponds to $V_{k} \simeq 120.66~ {\rm km/s}$ for $\eta \simeq 0.2257$ ($\alpha \simeq 0.5249$)}
\label{fitchett}
\end{center}
\end{figure}

 In Fig. (\ref{fitchett}) we plot the points $(V_{k},\eta)$ from Table 1. The continuous curve
is the least-square-fit of the points to the analytical formula
{\small
\begin{eqnarray}
\label{eq24}
V= A \eta^2 (1- 4 C \eta)^{1/2} (1+B \eta) \times 10^{3} ~{\rm km/s},
\end{eqnarray}}
with best fit parameters $A\simeq 3.63538199$, $B\simeq 2.46152380$ and $C \simeq 0.91379025$.
The maximum net antikick obtained is $\simeq 120.66 ~{\rm km/s}$ at $\eta \simeq 0.2257$.
Eq. (\ref{eq24}) is a modification of an empirical formula originally derived from
post-Newtonian analysis\cite{fitchett} and used by a number
of authors\cite{blanchet,gonzalez,blanchet1} to describe the distribution of kick velocities.
The additional parameter $C$ is necessary to account for the nonzero net gravitational
wave flux in non-head-on collision of two equal-mass black holes. Actually this modification introduced in (\ref{eq24}) is the only one
that works to produce an accurate fit, with rms error $\simeq 0.0376$ (and a normalized rms error ranging
from $0.5\%$ to $0.031\%$).
%
\begin{figure}
\begin{center}
{\includegraphics*[height=6.2cm,width=8.2cm]{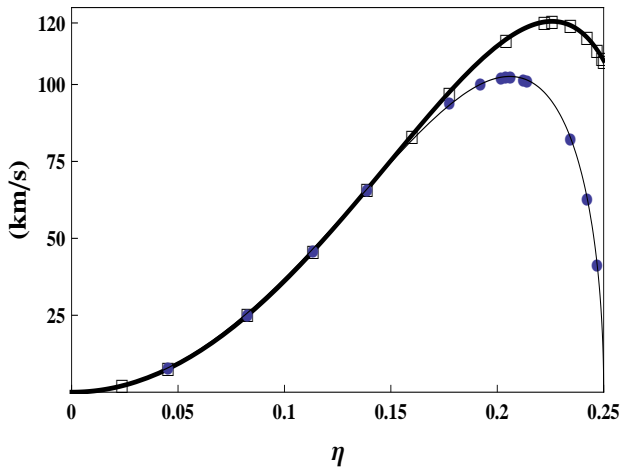}}
\caption{Comparison between distributions of the net kick velocities $V_k$ for a collision with incidence
angle $\rho_0=21^{o}$ (squares) and a head-on collision $\rho_0=0^{o}$ (circles). The head-on collision
is well fitted by a Fitchett $\eta$-scaling law ($C=1$) while the $\rho_0=21^{o}$ collision is fitted
by the modified Fitchett law with $C \sim 0.9138$ which accounts for the nonzero net
momentum flux for equal-mass black holes.}
\label{fitchett2}
\end{center}
\end{figure}
\par For comparison purposes we plot in Fig. \ref{fitchett2} the points $(V_k,\eta)$ for the case of a
head-on collision $\rho_0=0^{o}$\cite{aranha}, together with those of the non-head-on collision with $\rho_0=21^{o}$
given in Fig. \ref{fitchett}. The continuous curves are the best fit of the points to the analytical curve
(\ref{eq24}) with the best fit parameters $C=1$ for $(\rho_0=0^{o})$ (a genuine Fitchett curve) and
$C \simeq 0.91379025$ for $\rho_0=21^{o}$. As $\rho_0 \rightarrow 0$ the parameter $C \rightarrow 1$ monotonically,
as verified numerically. For the head-on case the distribution shows a consistent
similarity with the results of Refs. \cite{gonzalez,blanchet} where kicks in black hole inspiral binaries
are estimated from full numerical relativity simulations and from combining numerical relativity with perturbation theory.
This consistency is mainly connected to the order of magnitude of the kicks, and the location in $\eta$ of the maximum kick.
Using a Newtonian reasoning we may conclude that the momentum of each body in the inspiral binary
lies along the axis connecting the bodies and have the opposite direction, as seen from the center-of-mass
frame of the system. On this basis we might expect that the momentum fluxes in this system
and in a head-on collision present similar features.
\par We should note that, since our choice of $\gamma=0.5$ in the present paper was not fixed by any physical consideration,
a larger $\gamma$ in the initial data could be chosen so that the numerical values of the
maxima agree with, or have substantially larger values than the ones in \cite{gonzalez}.
For instance, for $\gamma=0.7$ ($v/c \simeq 0.6044$) and a value of $\alpha=0.524979$ (the latter corresponding to the
location of the maximum in the case $\gamma=0.5$, cf. Fig. \ref{fitchett}) we obtain $V_{k} \simeq 610.06 ~{\rm km/s}$,
with associated total net impulses $I_{W}^{x}(u_f)/m_0 \simeq -0.009753504379 $ and $I_{W}^{z}(u_f)/m_0 \simeq-0.016404424812$.
\par A remark is in order now, concerning the balance between the total rest mass of the remnant and the
total net impulse of the gravitational waves in the distribution of the net kick velocity
shown in Fig. \ref{fitchett}. As can be seen from Table 1, as $\eta$ increases from $0$ to $0.25$,
both the parameter $K_{f}$ and the total net impulse {\small $I_{W}(u_f)=\sqrt{I_{W}^{x}(u_f)^2+I_{W}^{z}(u_f)^2}$}
increase. However the increase of the rescaled Bondi rest mass of the remnant, $K_{f}^{3}$, is smaller than the increase of
$I_{W}(u_f)$ up to $\eta \simeq 0.225$, implying that in this range the net kick velocity increases in accordance
with (\ref{eq23}). Beyond this point the increase of $K_{f}^{3}$ is larger than the increase of the total
net impulse leading to a decrease in the values of $V_k$ up to $\eta=0.25$.
\par Finally we must comment that the parameter $C$ in the $\eta$-scaling law (\ref{eq24}), which assumes the value $C=1$
for head-on collisions $(\rho_0=0^{o})$ and for merging black hole inspirals, decreases monotonically as $\rho_0$ increases. In the case
$\rho_0=21^{o}$ we have $C \simeq 0.91379025$ as a best fit value while
for the limiting case $\rho_0=90^{o}$ we obtain $C \simeq 0.21165880$. In fact the quantity $(1-C)$, which
increases (from zero) as the incidence angle $\rho_0$ increases,
has a correspondence with the net kick velocity $V_{k}$, or equivalently with the nonzero total gravitational wave impulse $I_{W}(u_f)$,
for equal-mass colliding black holes ($\eta=0.25$). We mention that for the case $\rho_0=90^{o}$ the $\eta$-scaling law (\ref{eq24}) with
$C \simeq 0.21165880$ (and the other associated best fit values $A \simeq 1.94862882$ and $B \simeq 0.76602543$) fits
nicely the numerical points $(\eta,V_{k})$ obtained. However this curve increases monotonically with $\eta$,
a behavior analogous to the momentum distribution of the remnant black hole for $\rho_0=90^{o}$\cite{aranha11}.
These results will be the subject of a future publication.
\section{The center-of-mass motion of the merged binary system\label{section5}}

In our numerical work and its interpretation we were led to deal with several types of
velocity variables, as the net kick velocity, the initial infalling velocity of each individual
black hole, $v= \tanh \gamma$, and the final velocity of the remnant ${\bf v}_f=(\sin \rho_f,0,\cos \rho_f)~ v_f$.
Here we will further introduce the definition of an initial velocity ${\bf v}_{in}$ of the binary merged
system when the interaction of the two black holes and the initial gravitational wave content are taken into consideration.
Our main interest in the present Section is to examine the relation among all these variables and try to find
an appropriate definition of an approximate velocity of the center-of-mass of the merged system. The main problem in
this task is connected to the fact that the system loses mass due to the emission of
gravitational waves, which will lead us to look for a definition based on the momentum of
the radiation emitted and the associated conservation laws.
\par In Table 1 we have given the final velocity ${\bf v}_{f}$ for several $\eta$, which is directed
along the axis making an angle $\rho_f$ with the positive $z$-axis. By using Eqs. (\ref{eq21})-(\ref{eq22}) --
which are the components of the Bondi-Sachs integrated conservation law (\ref{eq18}) evaluated at $u=u_f$ --
we can relate ${\bf v}_f$ to the net kick velocity through the expression
{\small
\begin{eqnarray}
\label{eq25}
\frac{m_0 K_{f}^{3}}{c}~{\bf V}_{k}={\bf P}(u_f)-{\bf P}(u_0),
\end{eqnarray}}
where ${\bf P}(u_0)$ is the initial Bondi-Sachs momentum of the system which depends on $\alpha$, $\gamma$ and $\rho_0$,
and ${\bf P}(u_f)$ is the final momentum which depends on $\gamma_f$, $K_f$ and $\rho_f$.
The final velocity associated with the final momentum ${\bf P}(u_f)$ is given obviously by ${\bf v}_f$.
\par A tentative definition of the {\it initial velocity} of the merged system can be calculated from the
initial Bondi-Sachs momentum ${\small {\bf P}(u_0)}$ as ${\bf v}_{in}={\bf P}(u_0)/M_{B}(u_0)$,
where $M_{B}(u_0)$ is the initial Bondi mass of the system. ${\bf P}(u_0)$ can be evaluated
(with radiative corrections included) from the conservation law (\ref{eq18}) by taking $u=u_f$.
The modulus of ${\bf v}_{in}$ is obviously given by {\small $v_{in}=\sqrt{P^{x}(u_0)^2+P^{z}(u_0)^2}/M_{B}(u_0)$}.
The values of $v_{in}$, for several $\eta$, are listed in Table 1. We note that the initial velocity
$v_{in}$ is distinct from the previously defined initial infalling velocity ($v=\tanh \gamma$) in the following sense:
strictly speaking the interpretation of $v$ should actually be the initial infalling velocity of each hole at infinity before being
brought into interaction, when such quantity no longer has a separately defined meaning. On the contrary, $v_{in}$ is
the initial velocity at $u_0$ when the interaction of the two black holes and the initial gravitational wave content are already set up
in the initial data. For illustration, for $\gamma=0.5$ the initial infalling velocity is $v/c \simeq 0.4621$
while, for $\gamma=0.5$ and $\alpha=0.3$, say, we obtain $v_{in}/c\simeq 0.2874$ (corresponding to an effective $\gamma_{in} \simeq 0.2957$),
$v_{in}/c$ being of the order of magnitude of $v_f/c$.
\par
\begin{figure}[t]
\begin{center}
\vspace{0.7cm}
{\includegraphics*[height=6.0cm,width=8.2cm]{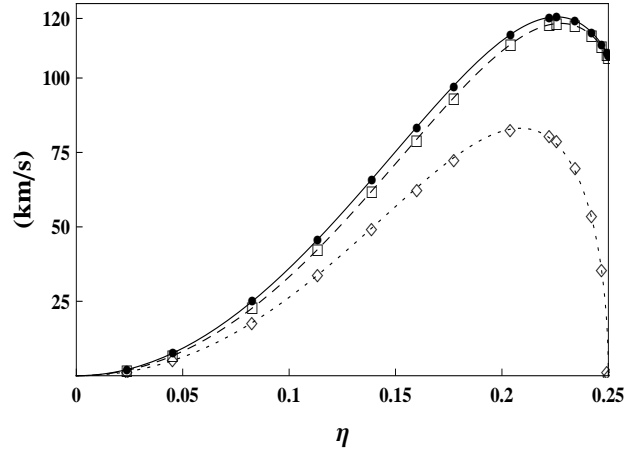}}
\caption{Plots of the points $\delta V =|{\bf v}_{in}-{\bf v}_{f}|$ (diamonds) and $\delta V_{CC}=(\zeta_0~{\bf v}_{in}-{\bf v}_{f})$ (squares) versus the
parameter $\eta$. The dotted line and dashed line are the least-square-fits of $\delta V$ and
$\delta V_{CC}$ to the formula (\ref{eq24}), respectively, with best fit parameters
($A\simeq 2.20190452$, $B\simeq 5.42498354$ and $C=1$) and ($A\simeq 3.08913213$, $B\simeq 3.47815851$ and $C \simeq 0.91126685$).
The maximum net velocity $\delta V_{CC}$ obtained is $\simeq 118.34~{\rm km/s}$ at $\eta \simeq 0.2277$.
The points (black circles) fitted by the solid line correspond to the net kick velocity of Fig. \ref{fitchett}, showing a
close approximation to the net velocity $\delta V_{CC}$.}
\label{figCMV}
\end{center}
\end{figure}

 A detailed examination of the numerical results in Table 1 shows that both $v_f$ and $v_{in}$ have a monotonic decrease
with $\eta$ (or equivalently with $\alpha$). Some features of the momentum extraction from the merged black hole by
gravitational waves -- connected, for instance, with the asymmetry of the radiation emitted (namely, with $\alpha$) and which are
present in $P_f$ and $V_{k}$ -- are erased in the distribution of $v_f$. This is not the case of
the difference of velocities defined by ${\bf \delta V}=({\bf v}_{f}-{\bf v}_{in})$, with magnitude
{\small
\begin{eqnarray}
\label{eq26}
\delta V = \sqrt{v_{in}^{2}+v_{f}^{2}-2~ {{\bf v}_{in}} \cdot {\bf v}_{f}},
\end{eqnarray}}
where ${{\bf v}_{in}} \cdot {\bf v}_{f}= v_{in} v_{f} \cos (\rho_{in}-\rho_f)$, and
$\rho_{in}=\arctan (P^{x}(u_0)/P^{z}(u_0))$.
The points ($\delta V, \eta$) have a distribution which is accurately fitted by
the $\eta$-scaling law (\ref{eq24}) with $C=1$ (cf. Fig. \ref{figCMV}),
with a maximum $\delta V \simeq 83.11 ~{\rm km/s}$ at $\eta \simeq 0.2087$.
\par This velocity variable $\delta V$ cannot however be related to the net kick velocity due to the
following problems. Not only $\delta V$ is much smaller than the net kick velocity (its maximum value being approximately
two thirds of the maximum value of the net kick velocity) but also its value is approximately zero
for $\alpha=1.0$ while the net kick velocity has a large value $V_{k} \simeq 107~{\rm km/s}$ at $\eta=0.25$.
Furthermore $v_{in}<v_f$ for all $\eta$ (cf. Table 1), which is not physically satisfactory since the
deceleration regime is dominant in the post-merger phase for all initial data parameters. The reason for these discrepancies is
that the velocity variable ${\bf \delta V}$ does not satisfy any momentum balance, while the net kick velocity
was evaluated from the momentum conservation law (\ref{eq18}).
\par Alternative choices that do not present such problems may be obtained from (\ref{eq18})
evaluated at $u=u_f$, which we express as
{\small
\begin{eqnarray}
\label{eq27}
{\bf v}_{in}-{\bf v}_{f}/\zeta_0=-~\frac{c~{\bf I}_{W}(u_f)}{M_B(u_0)},
\end{eqnarray}}
where {\small $\zeta_0 \equiv M_B(u_0)/(m_0 K_f^3 \cosh \gamma_f)$}.
\par Since in the nonlinear regime mass-energy is extracted by the emission of gravitational radiation,
we have obviously that the mass ratio $\zeta_0 > 1$, what -- in view of (\ref{eq27}) --  makes
the variable $\delta V$ inappropriate for a description of the dynamics (an exception being the nonphysical
scenario in which only momentum, and not mass, is extracted in the emission).
We are then led to consider appropriate net velocities constructed with ${\bf v}_{in}$ and ${\bf v}_f$ as
{\small
\begin{eqnarray}
\label{eq28}
{\bf \delta V}_C \equiv ({\bf v}_{in}-{\bf v}_{f}/\zeta_0) = - ~\frac{c~{\bf I}_{W}(u_f)}{M_B(u_0)},
\end{eqnarray}}
or
{\small
\begin{eqnarray}
\label{eq28-i}
{\bf \delta V}_{CC} \equiv  (\zeta_0~{\bf v}_{in}-{\bf v}_{f})= - ~\frac{c~{\bf I}_{W}(u_f)}{m_0 K_f^3 \cosh \gamma_f},
\end{eqnarray}}
where the mass ratio parameter {\small $\zeta_0$} corrects the definition $\delta V$ for the
loss of mass in the process.
\par The plots of the magnitudes $\delta V_{CC}$ and, for comparison, those of the net kick velocity $V_{k}$
and of $\delta V$ are exhibited in Fig. \ref{figCMV}. The points of $\delta V_{C}$ are not included
since the relative difference of its numerical values to the ones of
$\delta V_{CC}$ does not exceed $0.63\%$ for all $0< \eta < 0.25$, so the plots would not be distinguishable.
The dashed curve is the least-square-fit of the points ($\delta V_{CC},\eta$) (squares) to the formula (\ref{eq24}),
with best fit parameters $A\simeq 3.08913213$, $B\simeq 3.47815851$ and $C=0.91126685$. The maximum of $\delta V_{CC}$
obtained from the fit is $\simeq 118.34~{\rm km/s}$ at $\eta \simeq 0.2277$.
This distribution shows a close agreement with the distribution of the net kick velocity
(black circles and the solid curve). Finally the dotted curve is the least-square-fit of
the points ($\delta V, \eta$) (diamonds) to the law (\ref{eq24}) with $C=1$.
\par From Eq. (\ref{eq25}) we can then obtain an analytical relation
between the net kick velocity and $\delta V_{CC}$ and/or $\delta V_C$, given by
{\small
\begin{eqnarray}
\label{eq29}
{\bf V}_{k}=-~\cosh \gamma_f~ {\bf \delta V}_{CC}=-~\zeta_0~ \cosh \gamma_f~ {\bf \delta V}_C.
\end{eqnarray}}
It is worth commenting here that the close agreement of $\delta V_{CC}$ and/or $\delta V_{C}$ with the distribution of the
net kick velocity, as shown in Fig. \ref{figCMV}, is due to $\zeta_0$ being just slightly larger than $1$
for $0 < \eta < 0.25$.
For instance, for $\eta=0.0826$ ($\alpha=0.1$) we have $\zeta_0 \simeq 1.00033978$ and
for $\alpha=0.5$ we have $\zeta_0 \simeq 1.00317349$. This is sufficient to make $\delta V_C$ and/or $\delta V_{CC}$
positive and a few kilometers per second smaller than the net kick velocity. We also note that $\cosh \gamma_f \sim 1$
for the relevant domain of $\eta$  ($\gamma=0.5$ and $\rho_0=21^{o}$ fixed).
\par From the above results we are now in a position to suggest a possible definition
for the center-of-mass velocity of the merged binary. Our starting point is the fact that the velocity
of the center-of-mass of the remnant black hole is obviously ${\bf v}_f$, as measured by an inertial
observer at rest at infinity (cf. comment below Eq. \ref{eq5}). Also, by construction, the initial center-of-mass velocity of the merged
binary is approximately ${\bf v}_{in}$ and, for reasons discussed above, we adopt it as {\small $\zeta_0~ {\bf v}_{in}$}.
Relative to this same initial observer at infinity, and taking into account (\ref{eq27}), we are led to introduce
{\small
\begin{eqnarray}
\label{eq30}
{\bf v}_{\rm cm}(u) = \zeta_0 ~{\bf v}_{in}-{\bf \delta V}_{CC} (u) \equiv \zeta_0 ~{\bf v}_{in}+\frac{c~{\bf I}_{W}(u)}{m_0 K_f^3 \cosh \gamma_f},
\end{eqnarray}}
as a tentative definition of the center-of-mass velocity at any time $u \geq u_0$. For $u=u_f$ and $u=u_0$
we have by construction that ${\bf v}_{\rm cm}(u_f)={\bf v}_f$ and ${\bf v}(u_0)=\zeta_0 ~{\bf v}_{in} \sim {\bf v}_{in}$.
According to this definition, for an inertial frame ${\mathcal{L}}$ with velocity $\zeta_0 {\bf v}_{in}$
relative to a rest inertial frame at infinity, the motion of the center-of-mass is given by
{\small
\begin{eqnarray}
\label{eq31}
{\bf v}_{\rm cm}(u) \simeq -{\bf \delta V}_{CC} (u) \simeq \frac{c~{\bf I}_{W}(u)}{m_0 K_f^3 \cosh \gamma_f}.
\end{eqnarray}}
The final velocity of the remnant (in this inertial frame ${\mathcal{L}}$) is then
{\small
\begin{eqnarray}
\label{eq32}
{\bf v}_{\rm cm}(u_f)={\bf v}_f \simeq -~{\bf \delta V}_{CC}(u_f) \sim {\bf V}_{k}~,
\end{eqnarray}}
namely, the net kick velocity, as expected\footnote{We remark that the approximate equality signs in (\ref{eq31})-(\ref{eq33}) come from the
approximations made in the relativistic addition of velocities.}.
\par
\begin{figure}[t]
\begin{center}
{\includegraphics*[height=6.0cm,width=8.2cm]{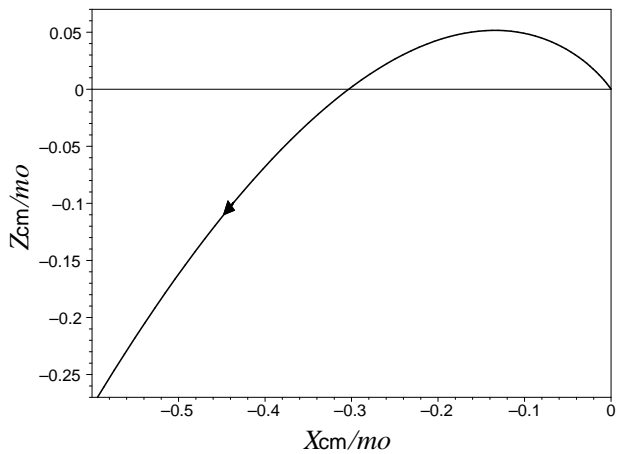}}
\caption{Plot of the trajectory of the center-of-mass of the binary merged system
($\alpha=0.2$ and $\rho_0=21^{o}$) in the inertial frame which is the rest frame of the center of mass at $u=u_0$.
The trajectory is an integral curve of the velocity field ${\bf v}_{\rm cm}(u)$, Eq. (\ref{eq33}),
with initial conditions $x_{\rm cm}=0=z_{\rm cm}$. The initial phase of positive impulse along the $z$-axis
is responsible for the initial inspiral branch of the trajectory in the $z>0$ semi-plane.
The trajectory has an asymptote as $u \rightarrow u_f$, which makes an angle $\Theta_f={\arctan (I_{W}^{x}(u_f)/I_{W}^{z}(u_f))}$
with the negative $z$-axis of the inertial frame ${\mathcal{L}}$ (cf. text). In the present case $\Theta_f \simeq 15.11^{o}$.}
\label{figCM}
\end{center}
\end{figure}

 We are now in a position to describe the motion of the center-of-mass of the merged binary system
in the inertial frame ${\mathcal{L}}$.
This inertial frame is actually the rest frame of the center of mass at $u=u_0$, and
the trajectory of the center-of-mass is an integral curve of the velocity field
{\small
\begin{eqnarray}
\label{eq33}
{\bf v}_{\rm cm}(u) \simeq \frac{c~{\bf I}_{W}(u)}{m_0 K_f^3 \cosh \gamma_f}.
\end{eqnarray}}
An integral curve of the velocity field (\ref{eq33}) with initial conditions $x_{\rm cm}=0=z_{\rm cm}$
is illustrated in Fig. \ref{figCM} for initial data parameters $\alpha=0.2$ and $\rho_0=21^{o}$.
This velocity field is proportional to the field $c~{\bf I}_W(u)=c~(I_{W}^{x}(u),0,I_{W}^{z}(u))$
so that the analysis of its integral curve can be discussed from the behavior of the curves in Fig. \ref{fig2}.
The positive impulse along the $z$-axis from $u=u_0$ to $u=u_{m}$ (resulting from the initial phase of
positive gravitational wave flux along $z$) is responsible for the inspiral branch
in the positive $z$ semi-plane. For $u \rightarrow u_f$ the curve
approaches the asymptote which makes an angle $\Theta_f={\arctan (I_{W}^{x}(u_f)/I_{W}^{z}(u_f))} \simeq 15.11^{o}$
with the negative $z$-axis of the inertial frame ${\mathcal{L}}$. This asymptote is the direction of
the net kick velocity, which is the velocity of the remnant in the inertial frame ${\mathcal{L}}$.
\section{Final Discussions and Conclusions}

The present paper extends and completes \cite{aranha11} by a detailed examination of the
gravitational wave recoil and the associated momentum extraction in the
post-merger phase of two black holes in non-head-on collision, in the realm of RT spacetimes.
The characteristic initial data for the system contain three
independent parameters, the mass ratio $\alpha$ of the two initial colliding black holes, the
initial infalling velocity parameter $\gamma$ and the incidence angle $\rho_0$.
Our analysis is based on the Bondi-Sachs momentum conservation laws which give a clear
picture of the processes that generate kicks in the binary merged system. The impulse imparted to the merged system by the
net momentum fluxes carried out by gravitational waves are evaluated as a function of time, for
a large domain of the initial data parameters. By numerically integrating the non-axisymmetric RT equation
(via a numerical code based on the Galerkin spectral method) we evaluate the total
net impulse at the final time $u_f$, when the gravitational wave emission ceases.
Typically the total net impulse is negative what, in terms of the Bondi-Sachs momentum conservation law,
corresponds to a dominant deceleration regime
of the system due to the emission of gravitational waves. However, for relatively small values of
the parameters $\alpha$ and $\rho_0$ an initial positive impulse in the $z$ direction may be present,
which will be responsible for an initial inspiral branch in the motion of the center-of-mass of the system.
A novel feature of non-head-on collisions ($\rho_0 \neq 0^{o}$) is the nonzero net gravitational wave fluxes
for equal-mass colliding black holes, contrary to the cases of head-on collisions and black hole inspiral binaries.
\par Based on the total net impulse we evaluated the net kick velocity\cite{gonzalez} imparted to the
system by the gravitational waves emitted. The distributions of the net kick velocity $V_k$ is
obtained as a function of the symmetric mass parameter $\eta$ and are
well fitted by a post-Newtonian motivated formula \cite{fitchett,blanchet} with an appropriate modification
to account for the nonzero gravitational wave momentum flux for equal-mass colliding black holes in
non-head-on collisions. The modified Fitchett $\eta$-scaling law adopted produces an accurate fit
of the points ($V_k,\eta$), with a rms error $\sim 0.038$. The maximum net kick obtained was $V_{k}\sim 121~{\rm km/s}$
located at $\eta \simeq 0.2257$, while for initial equal-mass black holes ($\eta=0.25$) we obtained $V_{k}\sim 107~{\rm km/s}$,
for an incidence angle $\rho_0=21^{o}$. Concerning the order of magnitude of the velocities and the location
of the maximum in $\eta$, these values are consistent with results of Refs. \cite{gonzalez,blanchet} which treated the case of
gravitational wave recoil in the merger of black hole binary inspirals. The larger values of the kick velocities
obtained in these references is possibly due to the fact that we have restricted our numerical integration to
a fixed initial data parameter $\gamma=0.5$ (or an initial infalling velocity $v/c \simeq 0.462$).
However we give numerical evidence that for larger $\gamma$'s the maximum kick velocity increases substantially.
In fact, for $\gamma=0.7$ and $\alpha=0.524979$ (the latter corresponding to the
location of the maximum in the case $\gamma=0.5$, cf. Fig. \ref{fitchett}) we obtain
$V_{k} \simeq 610.06 ~{\rm km/s}$.
\par The parameter $C$ in the modified Fitchett formula (\ref{eq24}), which assumes the value $C=1$ for head-on collisions
and binary black hole inspirals, decreases monotonically as the incidence angle $\rho_0$ increases.
Specifically the quantity $(1-C)$, which increases (from zero) as the incidence angle $\rho_0$ increases,
has a correspondence with the net kick velocity $V_{k}$ of equal-mass colliding black holes ($\eta=0.25$).
We mention that, for the limit case $\rho_0=90^{o}$, the points ($V_{k},\eta$)
are well fitted by (\ref{eq24}) with $C \simeq 0.2117$ but the fit curve is monotonic in $\eta$,
a behavior analogous to the momentum $P_f$ of the black hole remnant. A complete and detailed analysis of the
kick velocity distribution for a large domain of the initial data parameter space, with special focus on
its dependence on the incidence angle $\rho_0$, is currently under investigation.
.
\par Also based on the Bondi-Sachs momentum conservation law we have introduced a tentative definition
of the center-of-mass velocity of the binary merged system. In a inertial frame that closely approximates the
zero initial-Bondi-momentum frame, the center-of-mass velocity (i) is proportional to
the gravitational wave impulse vector ${\bf I}_{W}(u)$ (or the integrated momentum flux vector ${\bf P}_{W}(u)$); and
(ii) approaches the net kick velocity as $u \rightarrow u_f$. This inertial frame is  actually the rest frame
of the center-of-mass of the merged system at $u=u_0$ and the trajectory of the center-of-mass is an integral
curve of the velocity field ${\bf v}_{\rm cm} =c {\bf I}_{W}/ (m_0 K_{f}^{3} \cosh \gamma_f)$.
This integral curve exhibits an initial inspiral branch in the positive $z$ semi-plane whenever
an initial phase with $P_{W}^{z}(u)>0$ is present.
\par We must do a final comment on the connection between the simulations of the present paper
(post-common-horizon, head-on and non-head-on collisions in RT dynamics) and generic black hole merger kicks.
RT evolutions do not in general model the same physical situation as numerical relativity (NR) evolutions of merging binary inspirals,
and do not in general cover the same computational domain. In fact, although the physical process that generates recoils in RT
evolutions is the same as the one that generates recoils in NR evolutions,
namely, the balance between the variation of the momentum of the system and the momentum flux
of the gravitational wave emitted, the generation of the
momentum flux of the wave may be completely distinct in each case, depending on the
specification of the required initial data for the dynamics.
However the generated recoils in both approaches, specifically head-on collision of two black holes (RT)
and merging binary inspirals (NR), have impressive similar features. This suggests that
-- once a common horizon is formed -- the generation of the momentum flux of
the wave (which would determine the pattern of the recoils) might present in this domain a generic behavior
and that the characteristic approach can give complementary and reliable results on the momentum extraction of the system.
Also, as discussed in the present paper, characteristic initial data for
non-head-on collisions\cite{aranha11} satisfies a slightly modified Fitchett $\eta$-scaling law (\ref{eq24}), the modification
being necessary to account for the nonzero net momentum flux for equal-mass initial colliding black holes; the distribution
approaches the original Fitchett law as the incidence angle of collision approaches zero (the case of a head-on collision).
Recently the resulting recoil velocity associated with a family of parametrized initial data
(corresponding to a combination of the head-on collision characteristic initial data) was examined\cite{rezzolla1}
and was shown not to satisfy Fitchett's $\eta$-scaling law, although leading to a zero final recoil for equal-mass black holes.
\par The authors acknowledge the partial financial support of CNPq/MCTI-Brazil, through a Post-Doctoral Grant No. 201879/2010-7
(RFA), Research Grant No. 306527/2009-0 (IDS), and of FAPES-ES-Brazil (EVT). RFA acknowledges the hospitality and the partial financial
support of the Center for Relativistic Astrophysics, Georgia Institute of Technology, Atlanta, GA, USA, through a
Visiting Research Faculty-Exp grant, projects 4106680 and 4106C50.

\end{document}